\documentclass[aps,prl,twocolumn,showpacs]{revtex4}
\usepackage{amsfonts}
\usepackage{amsmath}
\usepackage{amsthm}
\usepackage{graphics}
\usepackage{graphicx}
\usepackage{subfigure}
\usepackage{amssymb}
\usepackage{graphics}
\usepackage{graphicx}
\usepackage{color}
\usepackage{subfigure}
\def\x{{\bf x}}

\def\kmax{k_{\rm max}}
\newcommand\bet{{g}}
 
\newcommand\alps{{\frac{\hbar^2}{2m}}}

\newcommand\dertt[1]{ \frac{\partial{ #1}}{\partial t} }
\newcommand\gd{\mbox{${\bf \nabla}^{2}$}}

\begin{document}

\title{Kelvin-wave cascade and dissipation in low-temperature superfluids vortices}

\author{Giorgio Krstulovic$^1$}
\affiliation{
Laboratoire Lagrange, UMR7293, Universit\'e de Nice Sophia-Antipolis, CNRS, Observatoire de la C\^ote d'Azur, B.P. 4229, 06304 Nice Cedex 4, France}

\pacs{67.25.dk, 47.37.+q, 67.25.dt, 03.75.Kk }

\begin{abstract}
We study the statistical properties of the Kelvin waves propagating along quantized superfluid vortices driven by the Gross-Pitaevskii equation. No artificial forcing or dissipation is added. Vortex positions are accurately tracked. This procedure directly allows us to obtain the Kevin-waves occupation-number spectrum. Numerical data obtained from long time integration and ensemble-average over initial conditions supports the spectrum proposed in [L'vov and Nazarenko, JETP Lett 91, 428 (2010)]. Kelvin wave modes in the inertial range are found to be Gaussian as expected by weak-turbulence predictions. Finally the dissipative range of the Kelvin-wave spectrum is studied. Strong non-Gaussian fluctuations are observed in this range.
\end{abstract}
\maketitle

Superfluid turbulence has been the subject of many experimental and theoretical works for the last decades. In particular a lot of progress has been done in experimental techniques and it is now possible to realize turbulent Bose-Einstein condensates (BEC) \cite{Henn2009}, turbulent flow with $^3$He \cite{Bradley2011,Eltsov2007} and visualize vortex filaments in $^4$He \cite{Bewley2006}. As in classical hydrodynamic turbulence \cite{Frisch1995}, a turbulent Kolmogorov cascade has been  observed experimentally and numerically. In superfluids, this takes place at scales larger than the mean inter-vortex distance $\ell$ \cite{Maurer1998,Nore1997a,Yepez2009}. At very low temperature, when damping due to mutual friction is negligible, it is believed that dissipation at small scales is carried by phonon radiation which dissipates energy into heat \cite{Vinen2002}. At scales smaller than $\ell$ the energy is transferred down by a series of reconnection processes of quantized vortices that excite waves on the filaments. These perturbations called Kelvin waves (KW), have been known for more than one century \cite{Thomson1880} in the broader context of fluid dynamics. These waves obey a set of non-linear equations where the energy transfer towards small scales is carried by a wave-turbulence cascade. How the energy is distributed along different scales is crucial for the understanding of the dissipative processes in superfluids. The energy spectrum of such a cascade is not yet fully-determined, except in the limit of small amplitude KW, where the theory of weak turbulence is applicable \cite{Zakharov1992}. However, a heated debate on the locality of KW energy transfer has taken place in the last years \cite{Laurie2010,Lebedev2010a,Kozik2010,Lebedev2010,Sonin2012,Boue2011}. Two different groups,  Kozik \& Stvistunov \cite{Kozik2004} and L'vov \& Nazarenko \cite{Lvov2010}, starting from the very same equations and by using the same theory, have derived two different spectra (hereafter KS and LN spectra repectivelly). The origin of this controversy is mainly due to a symmetry argument by KS (tilt of a vortex line) that eventually leads to a vanishing vertex in the perturbative expansion. This leads to locality in the energy transfer and makes the $6$-wave interaction theory realizable. The energy spectrum found by KS is
\begin{equation}
E_{\rm KS}(k)\sim \epsilon^{1/5} \kappa^{7/5}k^{-7/5}, \label{Eq:defE_KS}
\end{equation}
where $\epsilon$ is the energy flux, $\kappa$ the circulation quantum and $k$ the wavevector. This symmetry argument was questioned by LN and they claimed that the energy transfer is non-local. They derived an effective $4$-wave interaction theory that leads to the energy spectrum
\begin{equation}
E_{\rm LN}(k)\sim \kappa \epsilon^{1/3}\Psi^{-2/3}k^{-5/3}, \label{Eq:defE_LN}
\end{equation}
where $\Psi\sim (1/\kappa)\int E_{\rm LN}(k)dk$ is the mean-square angular deviation of the vortex from its straight-line configuration. More technical details on the controversy can be found in \cite{Lebedev2010a,Kozik2010,Lebedev2010,Sonin2012,Boue2011}. The exponent $7/5=1.4$ and $5/3\approx1.67$ of \eqref{Eq:defE_KS} and \eqref{Eq:defE_LN}, are respectively supposed to be universal, but their relatively close values makes it difficult to numerically elucidate which theory is correct. A number of numerical works supporting both theories have been published in the last years but none presenting strong arguments to settle this controversy \cite{Kozik2005,Boue2011,Baggaley2011}. These works are all done in the framework of the vortex filament (or equations derived from them) with an ad-hoc dissipative mechanism. It is worth mentioning that in the case of strong wave-turbulence, when the local slope of KW is order one, weak-turbulence breaks down and a different spectrum was proposed by Vinen et al. \cite{Vinen2003}. The Vinen spectrum $E_{\rm Vinen}(k)\sim k^{-1}$, was derived under the assumption of critical balance where the linear and non-linear time-scales of KW are of the same order. Finally, It was suggested by E. Sonin \cite{Sonin2012} that no universally can be expected for the KW spectrum. 

In this Letter, we address the small-amplitude KW cascade problem by performing direct numerical simulations of the Gross-Pitaevskii equation (GPE). The GPE formally describes the dynamics of a weakly-interacting BEC at very low temperature. It is also expected to at least qualitatively reproduce the dynamics of superfluid Helium vortices. As the Gross-Pitaevskii (GP) vortices can naturally radiate and excite phonons no artificial dissipation is needed, making it the natural framework for studying the KW cascade and the low-temperature dissipative processes. We first show that vortices coexist with small-scale thermalized phonon-waves.  Then, the ($1D$) KW occupation-number spectrum is precisely obtained and data is found to support the wave-turbulence prediction (LN) \cite{Lvov2010}. The KW spectrum is analyzed within the dissipative range and an exponential decay is found. Finally, the probability distribution function (PDF) of KW amplitudes is observed to be Gaussian in the inertial range in contrast with the power-law tails observed for modes in the dissipative range.

The GPE describing a homogeneous BEC of volume $V$ with  (complex) wave-function $\psi$ is given by
\begin{equation}
i\hbar\dertt{\psi} =- \alps \gd \psi + \bet|\psi|^2\psi ,
\label{Eq:GPE}
\end{equation}
where $m$ is the mass of the condensed particles and $g=4 \pi a \hbar^2 / m$, with $a$ the $s$-wave scattering length. Madelung's transformation $\psi({\bf x},t)=\sqrt{\frac{\rho({\bf x},t)}{m}}\exp{[i \frac{m}{\hbar}\phi({\bf x},t)]}$ relates the wave-function $\psi$ to a superfluid of density $\rho({\bf x},t)$ and velocity ${\bf v}={\bf \nabla} \phi$, where $\kappa=h/m$ is the Onsager-Feynman quantum of velocity circulation around the $\psi=0$ vortex lines. Equation \eqref{Eq:GPE} conserves the total energy $H=\int (\alps|\nabla \psi|^2 +\frac{g}{2}|\psi|^4)d\x$ and the total number of particles $N=\int |\psi|^2d\x$.
When Eq.\eqref{Eq:GPE} is linearized around a constant $\psi= \hat{\psi}_{\bf 0}$, the sound velocity is given by $c={(g| \hat{\psi}_{\bf 0}|^2/m)}^{1/2}$ with dispersive effects taking place at length scales smaller than the coherence length $\xi={(\hbar^2/2m|\hat{\psi}_{\bf 0}|^2g) }^{1/2}$ that also corresponds to the vortex core size.
In the GPE numerical simulations presented in this Letter the density $\rho=m N/V$ is fixed to $1$ and the physical constants in Eq.\eqref{Eq:GPE} are determined by the values of $\xi$ and $c=2$. The quantum of circulation $h/m$ has the value $c\,\xi/\sqrt{2}$. Numerical integration of Eq.\eqref{Eq:GPE} is performed by using a standard pseudo-spectral code with an exponential time-splitting temporal scheme in a cubic box of length $V^{1/3}=2\pi$. Resolutions used are listed in Table \ref{Table:RUNS} below, the value of $\xi$ is chosen as small as possible but well resolving vortex lines, and ensemble-averaging is done over $30$ initial conditions to reduce fluctuations.

\begin{figure}[htbp]
 \begin{center}
\includegraphics[height=2.5cm,width=0.87\columnwidth]{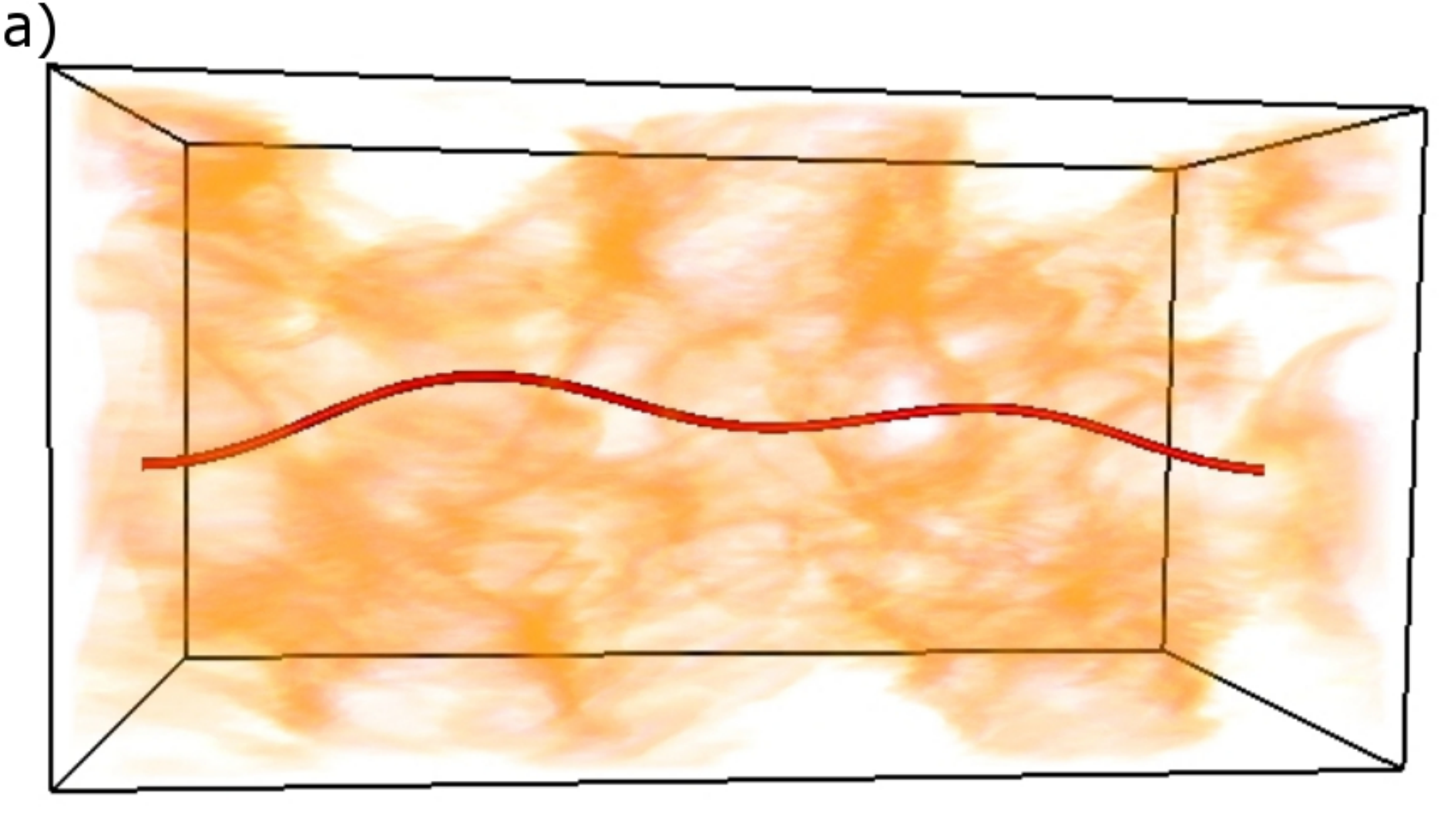}
\includegraphics[width=0.91\columnwidth]{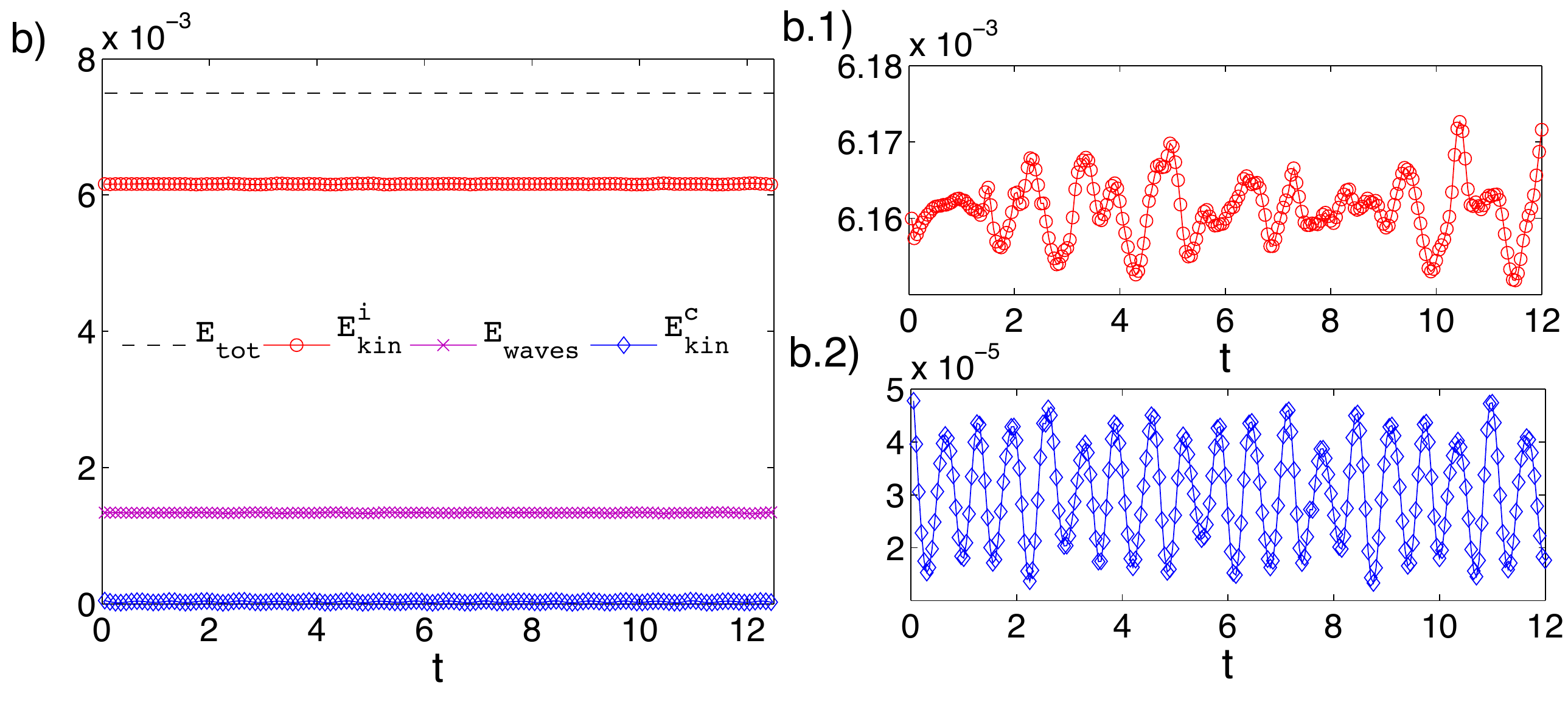}
\includegraphics[width=0.95\columnwidth]{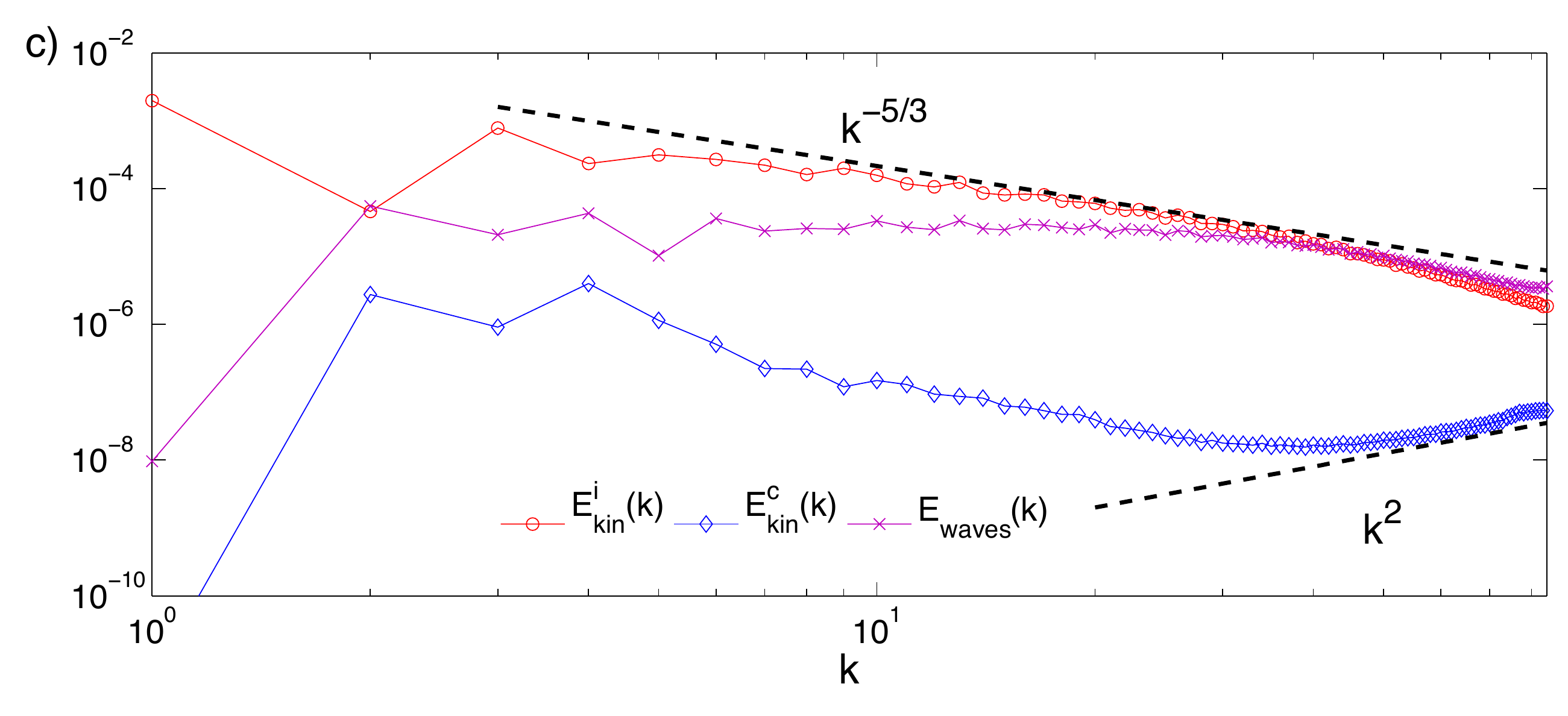}
\caption{a) (color online) 3D visualization of the density $|\psi|^2$ in the sub-box $[0,\pi]^2\times[0,2\pi]$.  In red an isosurface of the vortex and a (orange) density plot shows sound waves. b) Temporal evolution of total, incompressible kinetic, compressible kinetic and wave energy. b.1) and b.2) zoom of incompressible and compressible kinetic energy respectively. c) Incompressible kinetic, compressible kinetic and wave energy spectra. Dashed lines display $k^2$ and $k^{-5/3}$ power-law scalings.}
\label{Fig:fig1}
\end{center}
\end{figure}

To address the KW problem, we use the simplest configuration allowed by periodicity consisting of an array of four alternate-sign vortices. To obtain a clean initial condition and reduce the initial phonon emission, in a first step, an exact stationary solution of the GPE with straight vortices is numerically obtained by a Newton method \cite{Huepe:2000pp126-140}. Vortices are separated by a distance $\pi$ and can be considered isolated when $\xi\to0$, as the resolution is increased. Then, a KW is introduced slightly perturbing the vortex. The initial KW reads
\begin{equation}
x(z)=A\sum_n \cos{(n z+\phi^x_n)}\,, y(z)=A\sum_n \cos{(n z+\phi^y_n)},\label{Eq:IC}
\end{equation}
 where $\phi^{\{x,y\}}_n$ are random phases and $n=2,3$. A $3D$ visualization of the the density $\rho(\x)$ is displayed on Fig.\ref{Fig:fig1}a at $t=10$. 
The KW is clearly observed with the (red) isosurface. Phonon-waves correspond to the (orange) cloud that is a density-plot of $\rho(x)$ in a narrow threshold centered around the mean density value $\rho=1$. To quantify the vortical and wave energy of the configuration, we use the standard hydrodynamic energy decomposition, obtained by using the Madelung transformation (see \cite{Nore1997} for details). The total energy is thus decomposed in two terms: the incompressible kinetic energy $E_{\rm kin}^{i}$ containing the contribution of vortical structures and the energy of phonon-waves $E_{\rm wav}=E_{\rm kin}^{c}+E_{\rm int}+E_{\rm q}$, where $E_{\rm q}$,  $E_{\rm int}$, $E_{\rm kin}^{c}$,  are the quantum, internal and compressible kinetic energy respectively. Figure \ref{Fig:fig1}.b displays the temporal evolution of $E_{\rm kin}^{i}$, $E_{\rm kin}^{c}$, $E_{\rm wav}$ and $E_{\rm tot}=E_{\rm kin}^{i}+E_{\rm wav}$. Observe in Fig.\ref{Fig:fig1}b.1 and Fig.\ref{Fig:fig1}b.2 that their temporal evolution rapidly reach a (quasi-)statistical stationary regime. The same energy decomposition can be applied to the energy spectra that are displayed on Fig.\ref{Fig:fig1}c at $t=10$. The energy spectrum of the compressible kinetic energy presents at large wavenumbers a $k^2$-equipartition regime. This range is also present in the initial condition albeit with smaller values. It rapidly reaches the stationary state shown in Fig.\ref{Fig:fig1}c showing that thermalized waves coexist with vortices. We note that $E_{\rm kin}^{i}\ll E_{\rm wav}$, and hence the large-scale GPE dynamics is mainly driven by vortices setting a clean configuration to study the KW cascade problem.

The energy spectra displayed on Fig.\ref{Fig:fig1}.c present a scaling compatible with $k^{-5/3}$, however it cannot to be associated with Komogorov turbulence as the scale separation $V^{1/3}\gg \ell$ is not realized (here $V^{1/3}\sim\ell$ for all times). The scaling could be explained by the presence of a KW cascade and predictions \eqref{Eq:defE_KS} or \eqref{Eq:defE_LN}, as the principal contribution to energy of the fluid (see Fig.\ref{Fig:fig1}) is mainly due to vortices. However the relationship between the purely $1D$ ($D$ is the dimension) KW spectrum and $3D$ (hydrodynamical) energy spectra is not clearly stablished. To explicitly study the KW cascade, we numerically track the coordinates $(x(z),y(z))$ of the vortex. For each value of $z$ the equation $\psi(x(z),y(z))=0$ is solved by using a Newton method with the wave-function $\psi$ obtained from \eqref{Eq:GPE}. Derivatives of the fields at inter-mesh points are obtained by Fourier interpolation. This allows us to accurately obtain the vortex coordinates with a precision much larger than the one given by the mesh size or any other mesh-interpolation. Once the coordinates are obtained, it is possible to compute the KW vortex length, mean curvature and in particular the ($1D$) KW occupation-number spectrum (hereafter KW spectrum) defined by
\begin{equation}
n(k)=|\hat{w}(k)|^2+|\hat{w}(-k)|^2\label{Eq:defKelvonSpec},
\end{equation}
where $\hat{w}(k)$ is the Fourier transform of $w(z)=x(z)+iy(z)$. The KW spectrum allows us to directly construct the KW energy $E_{\rm KW}$ and the energy dissipation $\epsilon$:
\begin{equation}
E_{\rm KW}=\sum_k\omega(k)n_k\,,\hspace{.5cm}\epsilon=-\frac{dE_{\rm KW}}{dt}\label{Eq:defKelvonSpec}.
\end{equation}
where $\omega(k)$ is the KW dispersion relation. The KW dispersion relation can be approximated by $\omega(k)=C (\kappa/4\pi)k^2$, where $C$ is numerical constant, which eventually depends logarithmically  on $\xi/\ell$. 
As the previous quantities are computed from a $1D$ signal, they all present strong fluctuations. To reduce fluctuations, ensemble average is performed over the phases of the initial condition \eqref{Eq:IC}. 
 \begin{figure}[htbp]
 \begin{center}
\includegraphics[width=0.98\columnwidth]{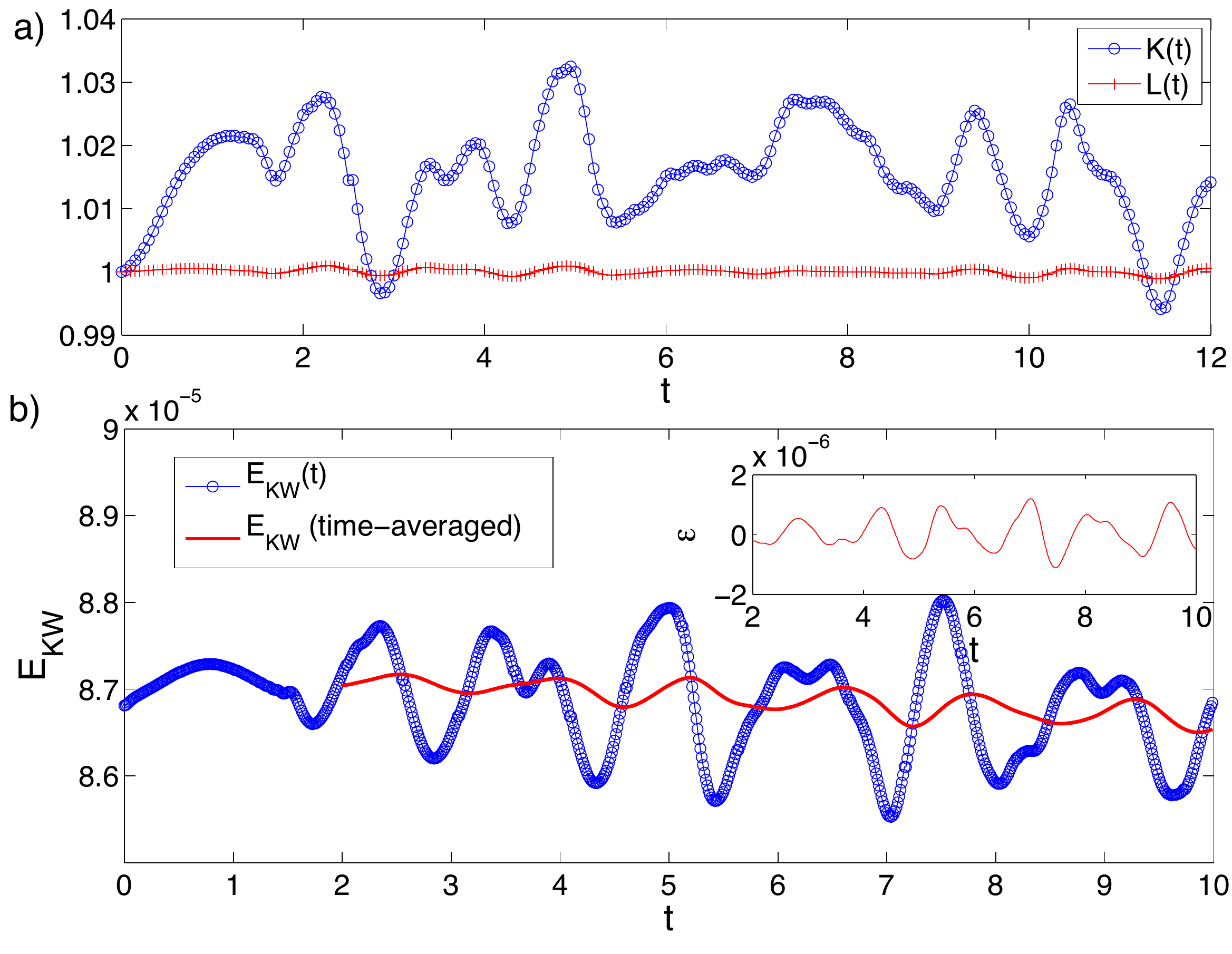}
\caption{(color online) a) Vortex length $L(t)$ (blue circles) and curvature $K(t)$ (red crosses) normalized by  $L(0)=6.51$ and $K(0)=0.44$. b) KW energy $E_{\rm KW}(t)$ (circles). The (red) solid line displays $E_{\rm KW}(t)$ averaged over temporal windows of width $\Delta t=2$. Inset: energy dissipation $\epsilon$. }
\label{Fig:fig2}
\end{center}
\end{figure}
Figures \ref{Fig:fig2}a and \ref{Fig:fig2}b show the temporal evolution of the total vortex length $L=\int\sqrt{1+|\partial_z w(z)|^2}dz$ and the mean curvature $K=\int|\partial_{z} w(z)\times \partial_{zz} w(z)|/|\partial_{z} w(z)|^3dz$, normalized by their initial values. Note that both quantities slightly fluctuate with time. Finally, on Fig.\ref{Fig:fig2}b the temporal evolution of the KW energy is displayed. Note that the energy fluctuates, especially after the arrival of phonon waves coming from neighboring cells near  $t\sim\pi/c\approx1.5$. The solid (red) line presents the energy averaged over temporal windows of width $\Delta t=2$; the decrease in energy is apparent. In the inset of Fig.\ref{Fig:fig2}b, the temporal evolution of the energy dissipation $\epsilon$ is shown. Note that the energy dissipation fluctuates and shows some negatives values as well. This can be related to the presences of phonon waves that excite KW at small scales.  However the temporal average is positive as more energy is radiated than absorbed by the vortex. The value of $\epsilon$ is not precisely determined here, as it depends on the constant $C$ and on the way temporal averaging is done, but in any case its mean value remains positive. 

We now turn to the KW spectrum. Two kinds of simulations are presented: The first, trying to enhance the scale separation between $V^{1/3}$ and $\xi$ and thus obtaining a larger inertial range. The second, concerns the dissipative range of the KW spectrum and then presenting a large number of modes between $\xi$ and smallest resolved scale $V^{1/3}/N_z$. The different runs and their details are listed on Table \ref{Table:RUNS}.

Let us focus now on the inertial range of the KW cascade. The two KS and LN wave-turbulence predictions for the KW spectrum read
\begin{equation}
n_{\rm KS}(k)=\frac{4\pi C_{\rm KS} \kappa^{2/5}\epsilon^{1/5}}{k^{17/5}},\hspace{.15cm}n_{\rm LN}(k)= \frac{4\pi C_{\rm LN}\epsilon^{1/3}}{\Psi^{2/3} k^{11/3}},\label{Eq:WTprediction}
\end{equation}
where $C_{\rm KS}$ and $C_{\rm LN}$ are numerical constant.

\begin{figure}[htbp]
 \begin{center}
\includegraphics[width=0.98\columnwidth]{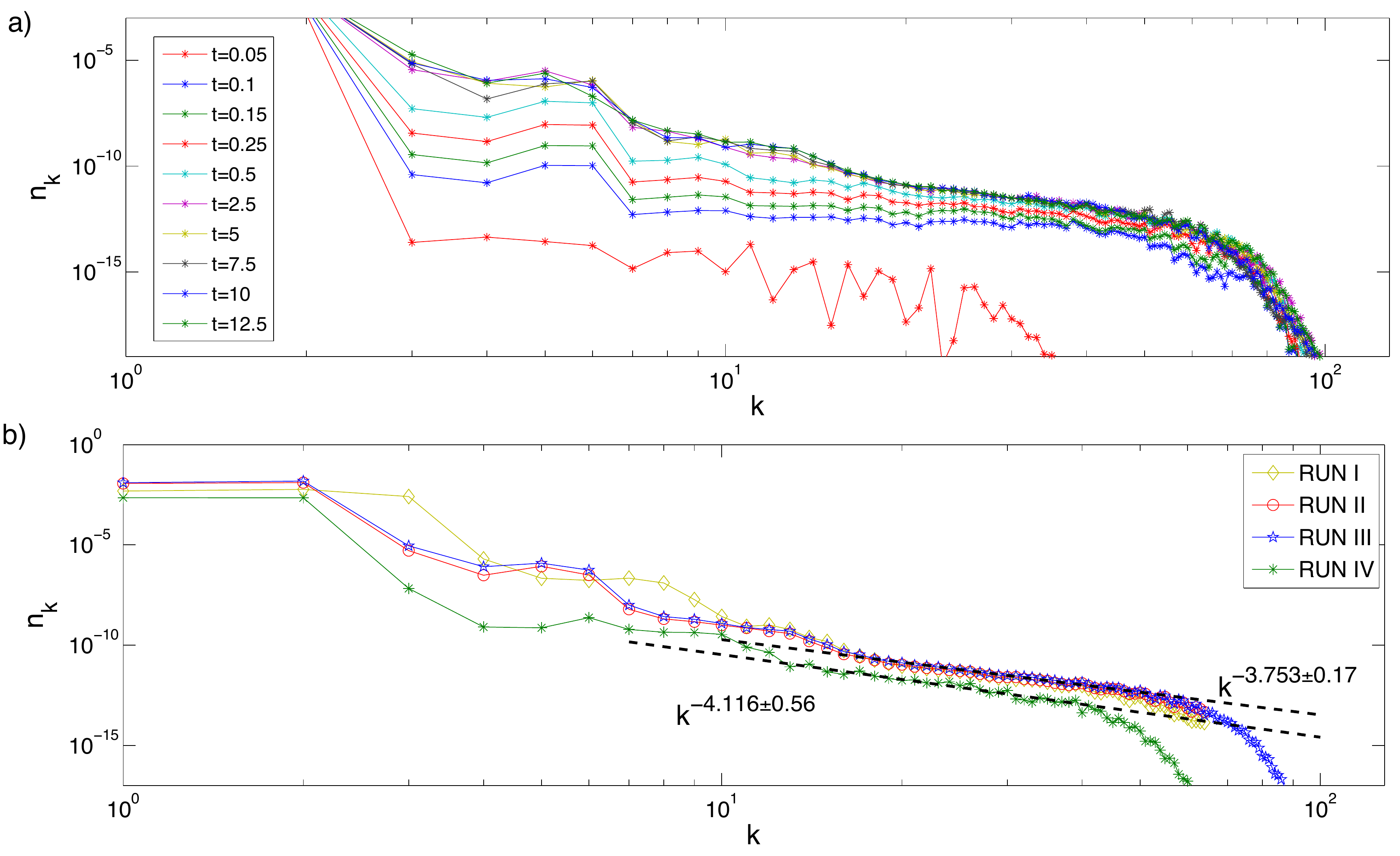}
\caption{a) Temporal evolution of KW spectrum, Run III. b) Time averaged KW spectra, RunS I-V. Dashed line displays the power-law fits.}
\label{Fig:fig3}
\end{center}
\end{figure}

The temporal evolution of the KW spectrum is displayed on Fig.\ref{Fig:fig3}.a for the Run III. 
KW are rapidly exited, and populate all wavenumbers. Energy arrives to scales small enough to be efficiently dissipated creating a steep decay zone usually called dissipative range in hydrodynamic turbulence \cite{Frisch1995}. As dissipation by phonon emission is very weak \cite{Krstulovic2008}, the system reaches a (quasi)-stationary state where a clear inertial range can be observed. To reduce fluctuations in the spectrum, time average during the stationary regime has also been performed. Fully averaged spectra of Runs I-IV are displayed on Fig.\ref{Fig:fig3}b. An inertial range with a power-law scaling is clearly appreciated for almost one decade. The exponent $m$, obtained from a fit $k^{-m}$, is shown on Table \ref{Table:RUNS} with their respective errors. 
\begin{table}[h]
\begin{tabular}{| c || c  |  c | c |c | c | c |  c |}
  \hline
  \hline
Run 	& $N_\perp$	& $N_z$	&$N_{\rm rea}$ 	&$n$ 	& $\xi$		& $A$	& $m$				 \\ \hline
I    	&  $256$ 		& $128$	&$31$			&$3$	& $0.025$		& $2\xi$ 	&   $3.85\pm0.24$			 \\
II	    	&  $256$ 		& $128$	& $31$			&$2$	&  $0.025$	&$4\xi$ 	&   $3.682\pm0.13$   		\\
III    		&     $256$ 	& $256$  	&$11$			&$2$	& $0.025$		& $4\xi$	& $3.753\pm0.17$					\\
IV     		&  $512$ 		& $256$	& $1$			&$2$	& $0.0125$	& $4\xi$	&$4.116\pm0.56$							\\
V     		&  $128$ 		& $512$	& $11$			&$2$	& $0.1$		& $4\xi$ 	&-													\\
     \hline
 \end{tabular}
 \caption{List of runs. $N_\perp$ and $N_z$ are the resolutions in the perpendicular and parallel directions respect to the vortex. $N_{\rm rea}$ is the number of realizations. $n$ corresponds to the number of initial KW modes and $m$ corresponds to the exponent $k^{-m}$ of the KW spectrum.\label{Table:RUNS}}
  \end{table}
Note that Run II and III have been performed with the same physical parameters but different resolutions in the $z$ direction resulting in almost identical spectra. The choice of resolution used in Run II allows us to speed up computations, and thus to get better statistics, at the expense of a reduced dissipative range. For all runs the exponent is slightly larger than the one predicted by the two weak-turbulence results \eqref{Eq:WTprediction} and presents a variation of a $5\%$. However, for all runs data supports the exponent $-5/3-2$ predicted by LN, that it is within errors, and excludes the $-7/5-2$ KS prediction. Experimentally, one usually have access only to the ($3D$) kinetic energy, that for small amplitude KW is dominated by the singularity of the velocity at the vortex core. A singularity cannot transfer energy and the KW cascade is thus crucial for understanding low-temperature dissipative-mechanisms of superfluids. The progress that is being achieved in visualization techniques of vortex filaments \cite{Bewley2006}, will maybe allow in the future to experimentally addressed this issue.

\begin{figure}[htbp]
 \begin{center}
\includegraphics[width=0.98\columnwidth]{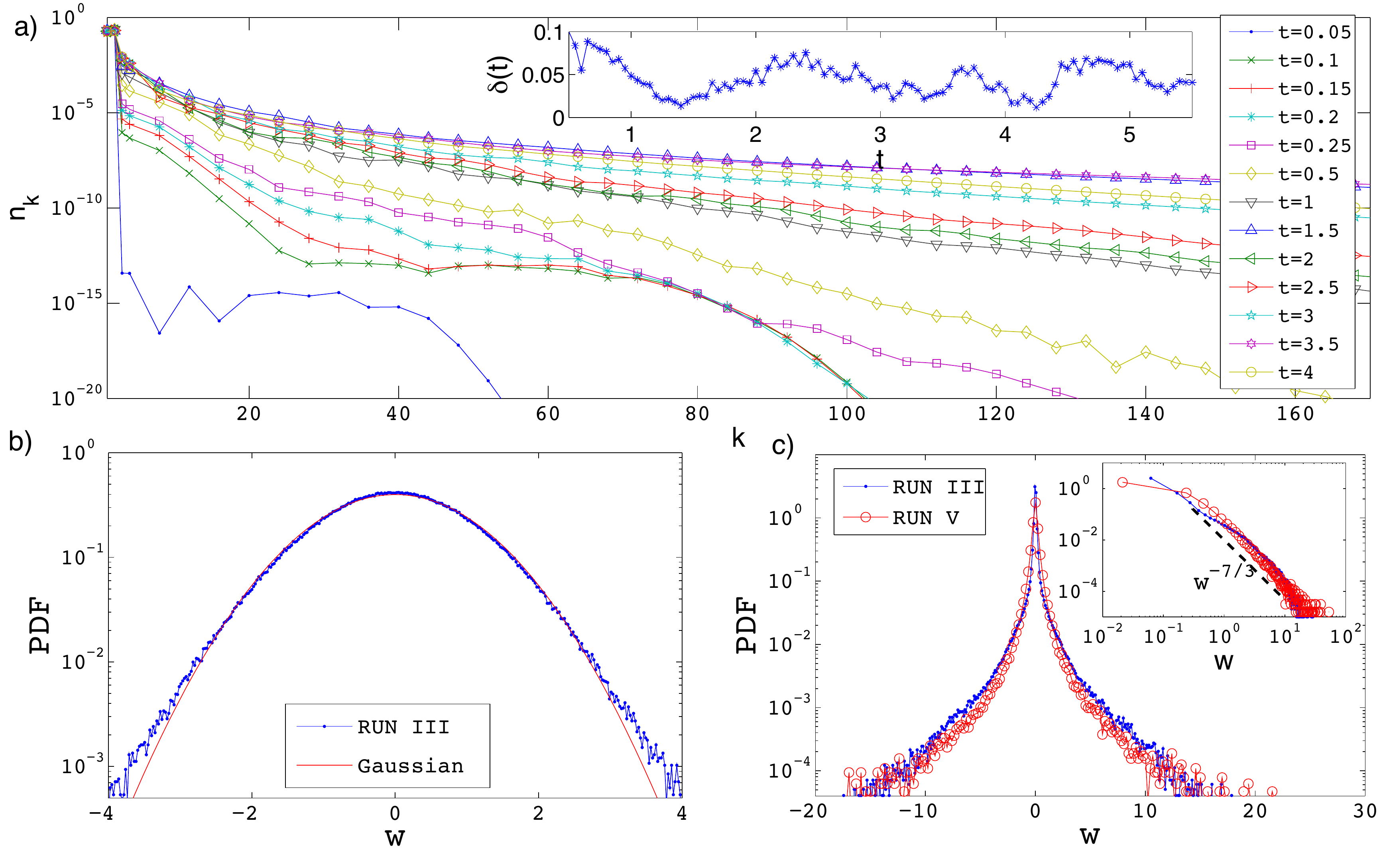}
\caption{ a) Temporal evolution of KW  spectrum for the Run V. Inset: temporal evolution of the exponential decay rate $\delta(t)$. b) PDF of KW amplitudes in the inertial range $20<k<40$ (Run III). c) PDF of KW amplitudes in the dissipative range $80<k<100$ (Run III) and $30<k<170$ (Run V). Inset: same PDF in Log-Log, the power-law $w^{-7/3}$ is drawn for reference}
\label{Fig:fig4}
\end{center}
\end{figure}

We now turn to the dissipative range of the KW spectrum, that takes place at scales much smaller than $1/\xi$. For such small scales it is known that dispersive effects of phonon waves slowdown the dynamics producing a bottleneck and quasi-thermalization \cite{Krstulovic2011a} of modes in an intermediate range. This effect was numerically observed for large values of $\xi\kmax$, where $\kmax$ is the largest wavenumber. A natural question is if this slowdown can affect the dissipative process of the KW cascade? If excitations of high wavenumber phonons are difficult, one could expect that dissipation of KW by sound emission should be reduced at such scales. Furthermore, at such small scales KW amplitudes should be somehow decoupled from the large-scale movement of the vortex and must be in direct interaction with out of equilibrium phonon waves. To investigate such a configuration we have performed simulations with a large value of $\xi\kmax=17$ (Run V). For such a configuration, the inertial-range of the KW cascade is not clearly identified in the KW spectrum presented in Fig.\ref{Fig:fig4}a.
A similar behavior to the one observed in the fully $3D$ dispersive bottleneck is observed at very early times. KW stop to be populated at  wavenumbers larger than $k_\xi=2\pi/\xi\sim60$. An equipartition regime of KW numbers for $k\lesssim k_\xi$ and faster than exponential decay for $k>k_\xi$ is observed. The similarity with the $3D$ dispersive bottleneck must be considered carefully. First, the dispersion relation of KW is dispersive at all wavenumber unlike the phonon dispersion relation (or Bogoliubov) where dispersive effects become important at scales smaller than $\xi$. Secondly, in the $3D$ dispersive bottleneck an equipartition of energy is observed and not equipartition of phonon occupation number. Furthermore, the statistics of the KW amplitudes in this range ($30<k<60$, $t=0.1$) is not Gaussian as expected in thermal equilibrium. At later times, the equipartition range is destroyed and a large-$k$ exponential decay is stablished showing the analyticity of the fields. The rate $\delta(t)$ (obtained with a fit $n_k\sim e^{-2\delta(t)k}$) is displayed in the inset of Fig.\ref{Fig:fig4}.a. It fluctuates in time with a minimum value $\delta_{\rm min}=0.007$. 

Finally we study the statistics of KW amplitudes in the inertial and dissipative range. The PDF of KW amplitudes varying at scales inside the inertial range for Run III is obtained by filtering in Fourier space and keeping modes between the range $20<k<40$. The normalized PDF is displayed in Fig.\ref{Fig:fig4}b together with a Gaussian distribution. The quasi-Gaussian behaviors (except for the very long tails) is manifest as expected from weak turbulence predictions. The square amplitude $|w(z)|^2$ consequently presents exponential tails. In the dissipative range (for  $80<k<100$) the PDF shows a strong non-Gaussian character as apparent on Fig.\ref{Fig:fig4}.c. Better statistics are obtained when considering Run V and modes between $30<k<170$. The PDF has power-law tails as shown on the inset of Fig.\ref{Fig:fig4}b.
Also, as is usual for turbulent flows \cite{Frisch1995}, an asymmetry is found when computing the skewness $\langle w^3\rangle/\langle w^2\rangle^{2/3}=-0.15$. Recently, in Biot-Savart simulations \cite{Baggaley2011a}, a crossover between Gaussian and non-Gaussian statistics was found in the velocity field taking place at the scale $\ell$. Here, for KW, the crossover takes place at the scale $\xi\ll\ell$. At scales smaller than $\xi$ the KW are somehow decoupled of the large scale dynamics prescribed by wave-turbulence and are in direct interaction with a strongly fluctuating superfluid velocity field, as the one found in the GP simulations of Ref. \cite{White2010}.

The behavior of KW at very small scales is an important issue, especially at very low temperature where dissipation by mutual friction is absent. In all vortex filament models, some small-scale artificial dissipation is needed to avoid energy pile-up at the smallest resolved scale. Although vortex-filament models are not concerned with such smalls scales, how the energy is dissipated in those models can affect the vortex dynamics and it would be important to check if the dissipative mechanisms used are consistent with the dissipation produced by phonon radiation in the GP results presented on Fig.\ref{Fig:fig4}.

To conclude, the KW cascade has been studied by performing Gross-Pitaevskii simulations of a slightly-perturbed straight vortex. Precise tracking of vortex coordinates and ensemble averaging allow us to the determination of a clean KW spectrum that supports the wave-turbulence prediction of L'vov and Nazarenko (LN) \cite{Lvov2010}. At scales smaller than the vortex size, an exponential decay of the KW spectrum was found. The PDF of KW was found to be quasi-Gaussian in the inertial range and to have power-law tails in the dissipative range. Precise tracking of GP vortices can be used in the future to improve the understanding of vortex thermal-waves interaction by using, for instance, the truncated (projected) GPE \cite{Krstulovic2011b}, where mutual friction effects are present.

\begin{acknowledgments}  
The author acknowledges useful scientific discussions with J. Bec, M.E. Brachet, J. Laurie, S.~Nazarenko, B.~Nowak, S.S. Ray and H. Salman. 
Computations were carried out at  M\'esocentre SIGAMM hosted at the Observatoire de la C\^ote d'Azur.
The research leading to these results has received funding from the European Research Council under the European CommunityÕs Seventh Framework Program (FP7/2007- 2013 Grant Agreement no. 240579).
\end{acknowledgments}


\end{document}